\documentclass[a4,12pt]{article}

\usepackage{graphicx}
\usepackage{amssymb}
\begin{document}

\begin{center}
{\large\bf Vanishing effective mass of the neutrinoless double beta
decay including light sterile neutrinos}
\end{center}

\vspace{0.3cm}

\begin{center}
{\bf Y.F. Li} $^{\rm a}$ \footnote{E-mail: liyufeng@ihep.ac.cn} ~
and ~ {\bf Si-shuo Liu} $^{\rm b}$
\footnote{E-mail: thingdx@mail.ustc.edu.cn} \\
{$^{\rm a}$ Institute of High Energy Physics, Chinese Academy of
Sciences, Beijing 100049, China \\
$^{\rm b}$ Department of Modern Physics, University of Science and
Technology of China, Hefei, Anhui 230026, China}
\end{center}

\setcounter{footnote}{0}

\vspace{2.5cm}

\begin{abstract}
Light sterile neutrinos with masses at the sub-eV or eV scale are
hinted by current experimental and cosmological data. Assuming the
Majorana nature of these hypothetical particles, we discuss their
effects in the neutrinoless double beta decay by exploring the
implications of a vanishing effective Majorana neutrino mass
$\langle m \rangle_{ee}$. Allowed ranges of neutrino masses, mixing
angles and Majorana CP-violating phases are illustrated in some
instructive cases for both normal and inverted mass hierarchies of three active neutrinos.\\

\hspace{-0.7cm} {PACS number(s): 14.60.Pq, 25.30.Pt, 14.60.St}
\end{abstract}

\newpage

\section{Introduction}

Despite the success of the standard three-neutrino oscillations in
explaining the results of solar, atmospheric, reactor and
accelerator neutrino experiments \cite{PDG} by two distinct
mass-squared differences, a number of anomalies from the short
base-line antineutrino experiments (e.g., LSND \cite{LSND} and
MiniBooNE \cite{Mini}) and the latest reactor antineutrino fluxes
\cite{R} indicate the existence of oscillations with much shorter
baselines. This would imply the existence of extra mass-squared
differences and therefore the mixing of active neutrinos with extra
sterile neutrino states \cite{Schwetz4,Giunti11a}. Furthermore, the
analysis \cite{Raffelt} of cosmic microwave background and large
scale structure data is hinting an existence of additional radiation
in the Universe, with sterile neutrinos being one of the plausible
candidates. Finally, one extra sterile neutrino is also allowed and
even favored from a recent analysis of the Big Bang Nucleosynthesis
bound \cite{Mangano}. Therefore, we should be open-minded on their
existence and it might be instructive to study the light sterile
neutrino hypothesis with other oscillation channels \cite{ice} or in
non-oscillation processes including the beta decay \cite{Riis} and
neutrinoless double beta decay \cite{Giunti10,review11}.

The most promising process to probe the Majorana nature of massive
neutrinos and acquire the information of Majorana CP-violating phases is the
neutrinoless double beta decay ($0\nu\beta\beta$)
\cite{review11,review03} of some even-even nuclei:
\begin{equation}
A(Z,N)\rightarrow A(Z+2,N-2)+2e^{-},
%     (1)
\end{equation}
which takes place via the exchange of Majorana neutrinos but would
be prohibited if neutrinos are the Dirac particles. The decay rate
is proportional to an effective Majorana neutrino mass term $\langle
m\rangle_{ee}$ and to the associated nuclear matrix element. The
latter can in principle be calculated with some uncertainties
\cite{NME} in nuclear physics. The effective neutrino mass $\langle
m\rangle_{ee}$ is related to the neutrino masses, mixing angles and
Majorana CP phases as follows:
\begin{equation}
\langle m\rangle_{ee}=\left|\sum_{i}m_{i}V_{ei}^2\right|,
%     (2)
\end{equation}
where $m_i$ is the mass of the $i^{th}$ neutrino mass state and the
sum of $i$ includes all light neutrino mass states. $V_{ei}$ stands
for an element in the first row of the lepton flavor mixing matrix,
and it contains a Majorana CP phase in general.

Great efforts have been made to search for signals of the
$0\nu\beta\beta$ process \cite{review11,review03} recently, however,
only negative results have been obtained and the upper bound on the
effective neutrino mass is achieved accordingly (i.e., $\langle
m\rangle_{ee}\lesssim 0.3-1.0\, \rm eV$ \cite{exps}). Although a
positive $0\nu\beta\beta$ signal would verify the Majorana nature of
massive neutrinos, negative results do not necessarily mean that
neutrinos are the Dirac particles. For Majorana neutrinos, a
vanishing effective neutrino mass $\langle m\rangle_{ee}$ due to
cancelation among different mass eigenstates could also lead to the
failure of $0\nu\beta\beta$ observations. This possibility in the
standard scenario of three-neutrino mixing has been explored in some
previous works \cite{xing03,merle}. If light (sub-eV) sterile
neutrinos exist and are mixed with their active counterparts, they
will also contribute to the $0\nu\beta\beta$ decay rate and modify
the prediction on the effective neutrino mass $\langle
m\rangle_{ee}$. Therefore, the parameter space of a vanishing
$\langle m\rangle_{ee}$ will get altered drastically in contrast to
the standard scenario. This point constitutes the main concern of
the present work. We shall derive the correlative relations among
neutrino masses, mixing angles and Majorana CP phases and constrain
the corresponding parameters by using current experimental data of
active and sterile neutrino oscillations.

The main part of this work is organized as follows. In section 2, we
shall describe the theoretical framework and derive the analytical
formulas for a vanishing $\langle m\rangle_{ee}$. Section 3 is
devoted to a numerical analysis. The analytical relations will be
confronted with current global oscillation data and numerical
examples will be illustrated in several specific cases. Finally, we
conclude in section 4.

\section{Analytical Calculations}

Assuming that neutrinos are the Majorana particles and there exist
$N^{}_{s}$ species of light sterile neutrinos, the neutrino flavor
eigenstates ($\nu_e$, $\nu_\mu$, $\nu_\tau$) are connected with the
$3+N^{}_{s}$ mass eigenstates ($\nu_i$, $i=1,\ldots,3+N_{s}$) by a
$3\times(3+N^{}_{s})$ flavor mixing matrix $V$ in the
charged-current interactions, which contains $3+3N^{}_{s}$ mixing
angles, $1+2N^{}_{s}$ Dirac CP phases and $2+N^{}_{s}$ Majorana CP
phases. In the $0\nu\beta\beta$ process, as revealed in Eq. (2),
only the mixing matrix elements in the first row of $V$ are relevant
in our calculations. Without loss of generality, we may redefine the
phases of three charged lepton fields in an appropriate way such
that the phase of the mixing matrix element $V^{}_{ei}$ is only of
the Majorana type. So the expression for the effective neutrino mass
$\langle m\rangle_{ee}$ can be rewritten as
\begin{eqnarray}
\langle
m\rangle_{ee}&=&\left|m_{1}|V_{e1}|^2e^{2i\rho^{}_{1}}+m_{2}|V_{e2}|^2e^{2i\rho^{}_{2}}+
m_{3}|V_{e3}|^2+\sum^{3+N^{}_{s}}_{j =
4}m_{j}|V_{ej}|^2e^{2i\rho^{}_{j}}\right|\,,
%     (3)
\end{eqnarray}
where $\rho^{}_{i}$ ($i=1,\ldots, 3+N^{}_{s}$) are the Majorana
phases and a vanishing $\rho^{}_{3}$ has been chosen. Although some
schemes with more than one sterile neutrino may have their distinct
properties such as CP violating effects in active-sterile neutrino
oscillations \cite{Schwetz4,Giunti11a}, we can combine all the
sterile neutrino contributions in the $0\nu\beta\beta$ process by
assuming an effective (3 + 1) scheme in which the additional
neutrino mass $m^{}_{0}$, mixing matrix element $|V^{}_{e0}|$ and
Majorana CP phase $\rho^{}_{0}$ are defined by
\begin{eqnarray}
m_{0}|V_{e0}|^2e^{2i\rho^{}_{0}}\equiv\sum^{3+N^{}_{s}}_{j =
4}m_{j}|V_{ej}|^2e^{2i\rho^{}_{j}}\,.
%     (4)
\end{eqnarray}
Taking account of Eq. (3) and (4), we find that a vanishing
effective neutrino mass (i.e.,$\langle m\rangle_{ee}=0$) requires
\begin{eqnarray}
m^{}_{0}|V_{e0}|^2\sin2\rho^{}_{0}+m^{}_{1}|V_{e1}|^2\sin2\rho^{}_{1}
+m^{}_{2}|V_{e2}|^2\sin2\rho^{}_{2}&=&0\,,\nonumber\\
m^{}_{0}|V_{e0}|^2\cos2\rho^{}_{0}+m^{}_{1}|V_{e1}|^2\cos2\rho^{}_{1}
+m^{}_{2}|V_{e2}|^2\cos2\rho^{}_{2}+m^{}_{3}|V_{e3}|^2&=&0\,.
%     (5)
\end{eqnarray}
Comparing these two conditions with current experimental data on the
mixing matrix elements ($|V^{}_{e1}|$, $|V^{}_{e2}|$, $|V^{}_{e3}|$
and $|V^{}_{e0}|$) and the mass-squared differences ($\Delta
m^{2}_{21}\equiv m^{2}_{2}-m^{2}_{1}$, $\Delta m^{2}_{31}\equiv
m^{2}_{3}-m^{2}_{1}$ and $\Delta m^{2}_{01}\equiv
m^{2}_{0}-m^{2}_{1}$), we might be allowed to determine or constrain
the absolute neutrino mass scale, neutrino mass hierarchies and
Majorana CP phases. Before doing the most general analysis, however,
let us discuss several specific cases to reveal the main
consequences of Eq. (5).

(a) The case of $m_{0}=0$ or $|V_{e0}|=0$ (or both) means that
contributions to $\langle m\rangle_{ee}$ are canceled out among
different sterile species of the mass eigenstates, which is
practically equivalent to the standard scenario of three-neutrino
mixing. Taking the (3 + 2) scheme with two additional sterile
neutrinos as an example, the cancelation between $\nu^{}_{4}$ and
$\nu^{}_{5}$ indicates
\begin{eqnarray}
\frac{m^{}_4}{m^{}_5}=-\frac{|V_{e5}|^2}{|V_{e4}|^2}\,,\quad |\,\rho^{}_{5}-\rho^{}_{4}|=\frac
{2n+1}{2}\pi\,,
%     (6)
\end{eqnarray}
with $n$ being an arbitrary non-negative integer. In this case, a
vanishing $\langle m\rangle_{ee}$ only requires the parameter
correlations among three active neutrinos,
\begin{eqnarray}
\frac{m_1}{m_2}=-\frac{|V_{e2}|^2}{|V_{e1}|^2}\frac{\sin{2\rho_2}}{\sin{2\rho_1}}
\,,\quad
\frac{m_2}{m_3}=+\frac{|V_{e3}|^2}{|V_{e2}|^2}\frac{\sin{2\rho_1}}{\sin{(2\rho_2-2\rho_1)}}\,,
%     (7)
\end{eqnarray}
which have already been studied in some earlier literatures
\cite{xing03,merle}. Therefore, we shall only discuss the
$0\nu\beta\beta$ process with non-vanishing sterile neutrino
contributions in the following parts of this work.

(b) The case of CP invariance requires that all the Majorana CP
phases should take some specific values (i.e.,
$\rho^{}_{i}={n_{i}\pi}/{2}$ with $n_{i}$ being arbitrary integers).
So the two conditions in Eq. (5) will be simplified to a single one,
\begin{eqnarray}
(-1)^{l_{0}}m^{}_{0}|V_{e0}|^2+(-1)^{l_{1}}m^{}_{1}|V_{e1}|^2+
(-1)^{l_{2}}m^{}_{2}|V_{e2}|^2+m^{}_{3}|V_{e3}|^2=0\,,
%     (8)
\end{eqnarray}
with $l_{k}=0\;{\rm or}\;1$ for ($k=0,1,2$), depending on $n^{}_{k}$
being even or odd integers respectively. Thus the mass spectrum of
active and sterile neutrinos might be fully determined for each
group of ($l_{0}$,$\,l_{1}$,$\,l_{2}$) with the help of current
knowledge on neutrino mass-squared differences and mixing matrix
elements. A simple calculation shows that (0, 1, 0) and (1, 0, 1)
are permitted for both mass hierarchies of active neutrinos, (1, 0,
0) and (0, 1, 1) are allowed only for the case of $\Delta
m^{2}_{31}>0$, but all the other possibilities are ruled out by
current global neutrino data.

(c) The case of a neutrino mass eigenstate being massless depends on
the sign of $\Delta m^{2}_{31}$. $m^{}_{1}=0$ or $m^{}_{3}=0$ might
hold for normal or inverted mass hierarchies respectively. If
$\Delta m^{2}_{31}>0$ and $m^{}_{1}=0$ hold, then Eq. (5) requires
\begin{eqnarray}
\frac{m_2}{m_0}=-\frac{|V_{e0}|^2}{|V_{e2}|^2}\frac{\sin{2\rho_0}}{\sin{2\rho_2}}
\,,\quad
\frac{m_3}{m_0}=+\frac{|V_{e0}|^2}{|V_{e3}|^2}\frac{\sin{(2\rho_0-2\rho_2)}}{\sin{2\rho_2}}\,;
%     (9)
\end{eqnarray}
on the other hand, in the case of $\Delta m^{2}_{31}<0$ and
$m^{}_{3}=0$, we are led to
\begin{eqnarray}
\frac{m_2}{m_0}=-\frac{|V_{e0}|^2}{|V_{e2}|^2}\frac{\sin{(2\rho_0-2\rho_1)}}{\sin{(2\rho_2-2\rho_1)}}
\,,\quad
\frac{m_1}{m_0}=+\frac{|V_{e0}|^2}{|V_{e1}|^2}\frac{\sin{(2\rho_0-2\rho_2)}}{\sin{(2\rho_2-2\rho_1)}}\,.
%     (10)
\end{eqnarray}
Since the masses of active neutrinos are totally calculable when we
assume $m^{}_{1}=0$ or $m^{}_{3}=0$, part of the parameter space of
Majorana CP phases must be excluded. Furthermore, the allowed region
of ($m^{}_{0}$, $|V_{e0}|$) might be constrained either by using Eq.
(9) or Eq. (10) from the active neutrino data or by using Eq. (4)
from the sterile neutrino data. Therefore, a comparison between the
two sectors may test the conditions of $\langle m\rangle_{ee}=0$ and
give complete constraints on the ($m^{}_{0}$, $|V_{e0}|$) parameter
space.

In addition to the special cases considered above, we can derive the
most general results about $(m_0,|V_{e0}|)$ and $\rho_0$ from the
conditions in Eq. (5). Indeed,
\begin{eqnarray}
&&m^{2}_{0}|V_{e0}|^4=m^{2}_{1}|V_{e1}|^4+m^{2}_{2}|V_{e2}|^4+m^{2}_{3}|V_{e3}|^4
+2m^{}_{1}m^{}_{2}|V_{e1}|^2|V_{e2}|^2\cos{(2\rho_1-2\rho_2)}
\nonumber \\
&&\quad\quad\quad\quad\quad
+2m^{}_{1}m^{}_{3}|V_{e1}|^2|V_{e3}|^2\cos{2\rho_1}
+2m^{}_{2}m^{}_{3}|V_{e2}|^2|V_{e3}|^2\cos{2\rho_2}\,,
%     (11)
\end{eqnarray}
as well as
\begin{equation}
\tan{2\rho_0}=\frac{m^{}_{1}|V_{e1}|^2\sin{2\rho_1}+m^{}_{2}|V_{e2}|^2\sin{2\rho_2}}
{m^{}_{1}|V_{e1}|^2\cos{2\rho_1}+m^{}_{2}|V_{e2}|^2\cos{2\rho_2}+m^{}_{3}|V_{e3}|^2}\,.
%     (12)
\end{equation}
Note from Eq. (11) and Eq. (12) that %,under the condition of $\langle m\rangle_{ee}=0$,
the parameters of sterile neutrinos (i.e., $m^{}_{0}|V_{e0}|^{2}$
and $\rho_0$) may be calculated from the active neutrino data,
depending on the absolute neutrino mass scale and neutrino mass
hierarchies. On the other hand, we may directly constrain
$m^{}_{0}|V_{e0}|^{2}$ and $\rho_0$ from the sterile neutrino data
by using Eq. (4). A combined analysis of both sectors will give the
complete parameter space allowed by the global neutrino data.

\section{Numerical Analysis}

To be explicit, let us carry out a numerical analysis of the
implications of $\langle m\rangle_{ee}=0$. For illustration, we
shall take the (3 + 2) neutrino mixing scheme with two sub-eV
sterile neutrinos as an example, which is more or less favored by
the latest experimental \cite{Schwetz4,Giunti11a} and cosmological
data \cite{Raffelt}. We may also assume that these sterile neutrinos
under consideration do not significantly affect the values of two
mass-squared differences and three mixing angles of active neutrinos
extracted from current active neutrino data \cite{Fogli,Schwetz3}.
In this assumption we shall use the latest results on active
neutrino parameters in Ref. \cite{Fogli} and sterile neutrino data
in Ref. \cite{Schwetz4}, namely, $\Delta m^2_{21} \approx 7.58
\times 10^{-5} ~{\rm eV}^2$, $\Delta m^2_{31} \approx 2.39\;(-2.31)
\times 10^{-3} ~{\rm eV}^2$, $\Delta m^2_{41} \approx 0.47 ~{\rm
eV}^2$ and $\Delta m^2_{51} \approx 0.87 ~{\rm eV}^2$ together with
$|V^{}_{e1}|\approx 0.824$, $|V^{}_{e2}|\approx 0.547$,
$|V^{}_{e3}|\approx 0.145$, $|V^{}_{e4}|\approx 0.128$ and
$|V^{}_{e5}|\approx  0.138$
%%%%%%%%%%%%%%%%%%%%%%%%
\footnote{An analysis of the latest global data has also been done
by a few independent groups, such as Ref. \cite{Schwetz3} for the
standard three-neutrino oscillations and Ref. \cite{Giunti11a} for
the active-sterile neutrino oscillations .}
%%%%%%%%%%%%%%%%%%%%%%%%%
as typical inputs of the best-fit values. The sign of $\Delta
m^2_{31}$ is unknown and the absolute neutrino mass scale, which can
be characterized by the smallest neutrino mass ($m^{}_{1}$ for
$\Delta m^2_{31}>0$  or $m^{}_{3}$ for $\Delta m^2_{31}<0$), is
expected to be at most of ${\cal O}(0.1)$ eV or ${\cal O}(1)$ eV as
constrained by current experimental and cosmological data
\cite{PDG,WMAP10}. On the other hand, the Majorana CP phases are
totally unconstrained and will vary in the full parameter space.
Finally, we shall take into account the uncertainties of active
neutrino parameters at the $1\sigma$ or $3\sigma$ confident level in
our numerical analysis but neglect the uncertainties of sterile
neutrino parameters which are not explicitly presented in Ref.
\cite{Schwetz4}. We admit that our calculations are quite
preliminary and mainly serve for illustration, but it should be
emphasized that they indeed reveal the main consequences of $\langle
m\rangle_{ee}=0$.

Now we are able to do the numerical analysis on possible
implications of $\langle m\rangle_{ee}=0$. We shall employ the
analytical results of different cases discussed in section 2 as our
guidelines and plot the corresponding regions allowed by the global
oscillation data.

In the cases of the smallest neutrino mass being zero, it is
straightforward to present the correlative relations among different
Majorana CP phases and between ($m^{}_{0}$, $|V_{e0}|$). First we
illustrate the case of $m^{}_{1}=0$ in Fig. 1. The parameter space
of two Majorana phases ($\rho^{}_{0}$, $\rho^{}_{2}$) is shown in
Fig. 1(a), where different (black and grey) colors stand for
$1\sigma$ and $3\sigma$ ranges of the active neutrino parameters
respectively. One should note that $\rho^{}_{1}$ is irrelevant in
this case because the massless mass eigenstate $\nu_1$ does not
contribute to the $0\nu\beta\beta$ process. In Fig 1(a), we can
learn that there are two isolated regions to guarantee $\langle
m\rangle_{ee}=0$, and they are related through the transformation of
($\rho^{}_{0}\rightarrow\pi-\rho^{}_{0}$, $\rho^{}_{2}\rightarrow
\pi-\rho^{}_{2}$). This point can be understood by considering the
fact that the formulas in Eq. (9) are invariant under the same
transformation. Furthermore, the allowed regions can be generalized
to the larger ($\rho^{}_{0}+ n_{0}\pi$, $\rho^{}_{2}+ n_{2}\pi$)
parameter space, where $n_{0}$ and $n_{2}$ are arbitrary integers.
The allowed regions of ($m^{}_{0}$, $|V_{e0}|$) can be found in Fig.
1(b), where the (black and grey) shaded regions stand for
constraints from the active neutrino data and the region with (red)
sparse lines is required by the sterile neutrino parameters. The
(black) horizontal line stands for the $2 \sigma$ upper bound on the
sum of the neutrino masses from cosmological probes \cite{cosmosnu}
and only the region below this line is allowed. The profiles of
these allowed regions have the shape of the inverse-rate curve in
the single logarithmic scale, which can be easily understood by the
analytical expressions in Eq. (4) and Eq. (5). One should also note
that there is no overlap between the allowed regions of active and
sterile neutrino data even within the $3\sigma$ ranges, indicating
that the case of $m^{}_{1}=0$ and $\Delta m^2_{31}>0$ is disfavored
by current global oscillation data if $\langle m\rangle_{ee}=0$
holds. This conclusion demonstrates the fact that the minimum of
$m^{}_{0}|V_{e0}|^{2}$ from the sterile neutrino data,
$|\sqrt{\Delta m^2_{41}}|V_{e4}|^{2}-\sqrt{\Delta
m^2_{51}}|V_{e5}|^{2}|$ is always larger than the maximum of
$m^{}_{0}|V_{e0}|^{2}$ from the active neutrino data, $|\sqrt{\Delta
m^2_{21}}|V_{e2}|^{2}+\sqrt{\Delta m^2_{31}}|V_{e3}|^{2}|$.

A similar analysis for $\Delta m^2_{31}<0$ and $m^{}_{3}=0$ is given
in Fig. 2. The allowed regions of Majorana CP phases and
($m^{}_{0}$, $|V_{e0}|$) are illustrated in the upper and lower
panels respectively. Only the differences of Majorana phases can be
constrained because it is the third mass eigenstate $\nu_3$ that
decouples from the effective mass $\langle m\rangle_{ee}$ due to its
masslessness. In Fig. 2(a), we take ($\rho^{}_{0}-\rho^{}_{1}$) and
($\rho^{}_{2}-\rho^{}_{1}$) as free parameters and find that they
are strongly correlated and constrained. The area of the allowed
region here is much smaller than that in Fig. 1(a) simply for the
reason that the mixing matrix elements of active neutrinos in Eq.
(10) (i.e., $|V_{e1}|$, $|V_{e2}|$) are much more precise than that
appearing only in Eq. (9) (i.e., $|V_{e3}|$). Now let us take a look
at Fig. 2(b) for the allowed regions of ($m^{}_{0}$, $|V_{e0}|$).
The (black) horizontal line stands for the $2 \sigma$ upper bound on
the sum of the neutrino masses from cosmological probes
\cite{cosmosnu}. Its most distinct difference from Fig. 1(b) is the
overlaps between active and sterile neutrino constraints appearing
at both $1\sigma$ and $3\sigma$ levels. Therefore, $m^{}_{3}=0$ and
$\langle m\rangle_{ee}=0$ can be simultaneously satisfied in the
presence of sub-eV sterile neutrinos. In the meantime, it is
instructive to compare the present scenario with the standard
scenario of three-neutrino mixing, where a vanishing $\langle
m\rangle_{ee}$ favors $\Delta m^2_{31}>0$ over $\Delta m^2_{31}<0$.

So far we have only discussed the consequences of a vanishing
$\langle m\rangle_{ee}$ in some specific cases. To illustrate the
dependence on the absolute neutrino mass scale, we calculate the
allowed ranges of $m^{}_{0}|V_{e0}|^{2}$ versus the smallest
neutrino mass ($m^{}_{1}$ or $m^{}_{3}$) for both neutrino mass
hierarchies in Fig. 3. We give the constraints from the active
neutrino data with (black and grey) shaded regions and constraints
from the sterile neutrino data with regions of (red) sparse lines.
The overlapping regions of both sectors are the complete parameter
space of $\langle m\rangle_{ee}=0$ and the global oscillation data.
The case of $\Delta m^2_{31}>0$ is illustrated in Fig. 3(a), which
confirms the previous conclusion that the specific case of
$m^{}_{1}=0$ is disfavored by current global oscillation data. As
$m^{}_{1}$ is getting larger, the active neutrino spectrum changes
from the hierarchical type to the quasi-degenerate one and therefore
the relations between $m^{}_{0}|V_{e0}|^{2}$ and $m^{}_{1}$ in Eq.
(11) approximate to linear functions. Numerically, we may find the
allowed ranges explicitly as $0.005~{\rm eV}\lesssim
m^{}_{1}\lesssim 0.116~{\rm eV}$ and $0.006~{\rm eV}\lesssim
m^{}_{0}|V_{e0}|^{2}\lesssim 0.029~{\rm eV}$. In comparison, a
similar calculation for $\Delta m^2_{31}<0$ can be found in Fig.
3(b) and the ranges of $m^{}_{3}$ and $m^{}_{0}|V_{e0}|^{2}$ are
$0~\lesssim m^{}_{3}\lesssim 0.10~{\rm eV}$ and $0.012~{\rm
eV}\lesssim m^{}_{0}|V_{e0}|^{2}\lesssim 0.029~{\rm eV}$,
respectively. Finally, let us comment on the Majorana CP phase
$\rho^{}_{0}$ in Eq. (12). Although it can be calculated from the
active neutrino data, there is no correlation between $\rho^{}_{0}$
and the absolute mass scale (characterized by $m^{}_{1}$ or
$m^{}_{3}$). Our numerical calculation shows that $\rho^{}_{0}$ may
run over the full parameter space at each point of the smallest
neutrino mass ($m^{}_{1}$ or $m^{}_{3}$). Only when the neutrino
mass scale is fixed, as the specific cases discussed above, there
can be non-trivial correlations among different Majorana CP phases.
Within the full parameter ranges, it is instructive to estimate the
minimal value of the effective Majorana mass (i.e., $\langle
m\rangle^{min}_{ee}$) where $\langle m\rangle_{ee}$ does not vanish.
We calculate $\langle m\rangle^{min}_{ee}$ versus the smallest
neutrino mass ($m^{}_{1}$ or $m^{}_{3}$) for both $\Delta
m^2_{31}>0$ and $\Delta m^2_{31}<0$ cases in Fig. 4. The black solid
and grey dashed lines stand for $1\sigma$ and $3\sigma$ ranges of
the active neutrino data respectively. When $\Delta m^2_{31}>0$
holds, there are two separated ranges of $m_1$ for a non-vanishing
$\langle m\rangle^{min}_{ee}$. A hierarchical neutrino mass spectrum
with $m_1\lesssim 10^{-3}~{\rm eV}$ predicts a minimal value of
$\langle m\rangle_{ee}$ as $(3\sim4)\times10^{-3}~{\rm eV}$ and the
magnitude of $\langle m\rangle^{min}_{ee}$ can reach the ${\cal
O}(0.1)~{\rm eV}$ level for a nearly degenerate spectrum of the
active neutrinos. In contrast, a non-vanishing $\langle
m\rangle^{min}_{ee}$ can only achieved for $m_3\gtrsim 0.1~{\rm eV}$
in the case of $\Delta m^2_{31}<0$ and $\langle m\rangle^{min}_{ee}$
grows as $m_3$ becomes larger.

\section{Concluding Remarks}

Motivated by non-observation of the $0\nu\beta\beta$ process and
possible hints of sub-eV sterile neutrinos from the latest
experimental and cosmological data, we have discussed the
implications of a vanishing effective Majorana mass (i.e., $\langle
m\rangle_{ee}=0$) in the presence of light sterile neutrinos.
Current neutrino oscillation data from both active and sterile
neutrinos do allow $\langle m\rangle_{ee}=0$ to hold if neutrinos
are the Majorana particles and their masses, mixing angles and
Majorana CP phases possess some specific correlations. The allowed
parameter space can be drastically altered in contrast to that of
the standard three-neutrino mixing scenario. For the standard
scenario, a vanishing $\langle m\rangle_{ee}$ favors $\Delta
m^2_{31}>0$ over $\Delta m^2_{31}<0$. However, it is not really the
case after we include the light sterile neutrino contributions. In
the specific case where the smallest neutrino mass vanishes, a
vanishing $\langle m\rangle_{ee}$ is consistent with the case of
$\Delta m^2_{31}<0$ rather than $\Delta m^2_{31}>0$. Furthermore,
for the complete parameter ranges, both cases are allowed and we may
constrain the smallest neutrino mass as $0.005~{\rm eV}\lesssim
m^{}_{1}\lesssim 0.116~{\rm eV}$ and $0~\lesssim m^{}_{3}\lesssim
0.10~{\rm eV}$ respectively.

Going beyond the Standard Model, there are different mechanisms
\cite{Probe} that may be able to mediate the $0\nu\beta\beta$
process, including light (active and sterile) Majorana neutrinos,
heavy sterile Majorana neutrinos \cite{xing09,RHN} and other
particles \cite{LR} that can induce the lepton-number-violating
(LNV) interactions. We stress that we have simply neglected all the
contributions other than light Majorana neutrinos and our
conclusions will be invalid once other mechanisms can give a
comparable decay rate as either light active neutrinos or light
sterile neutrinos. In addition, when more accurate data on neutrino
mass and mixing parameters and more sensitive observations or
constraints on the $0\nu\beta\beta$ process are achieved in the
future \cite{Futu11,Futu22}, it might be possible to confirm or rule
out the parameter space we have obtained and even probe the
existence of light sterile neutrino states.

\vspace{0.5cm} We would like to thank Prof. Zhi-zhong Xing for
suggesting this topic and very helpful discussions. One of us (SXL)
is also indebted to him for the warm hospitality and financial
support at the theoretical physics division, Institute of High
Energy Physics in Beijing, where this work was done. YFL is
supported by the China Postdoctoral Science Foundation under grant
No. 20100480025.

\newpage

%%%%%%%%%%%%%%%%%%%% Fig. 1 %%%%%%%%%%%%%%%%%%%%%%%%%%%%%%
\begin{figure}[p!]
\begin{center}
\begin{tabular}{cc}
\includegraphics*[bb=24 16 292 236, width=0.70\textwidth]{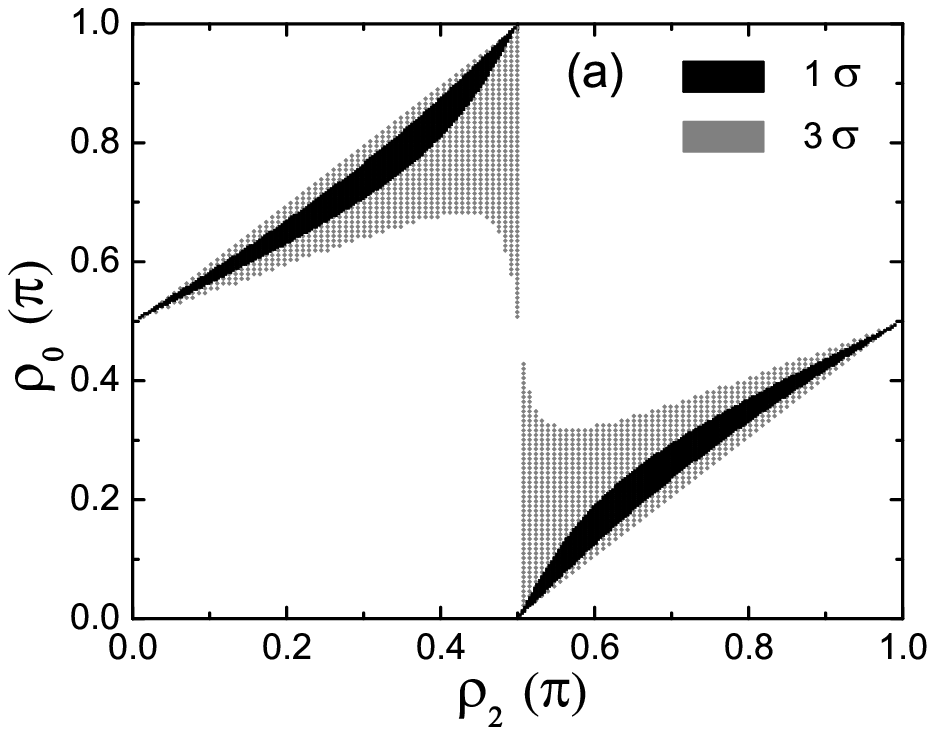}
\\
\includegraphics*[bb=22 20 290 240, width=0.70\textwidth]{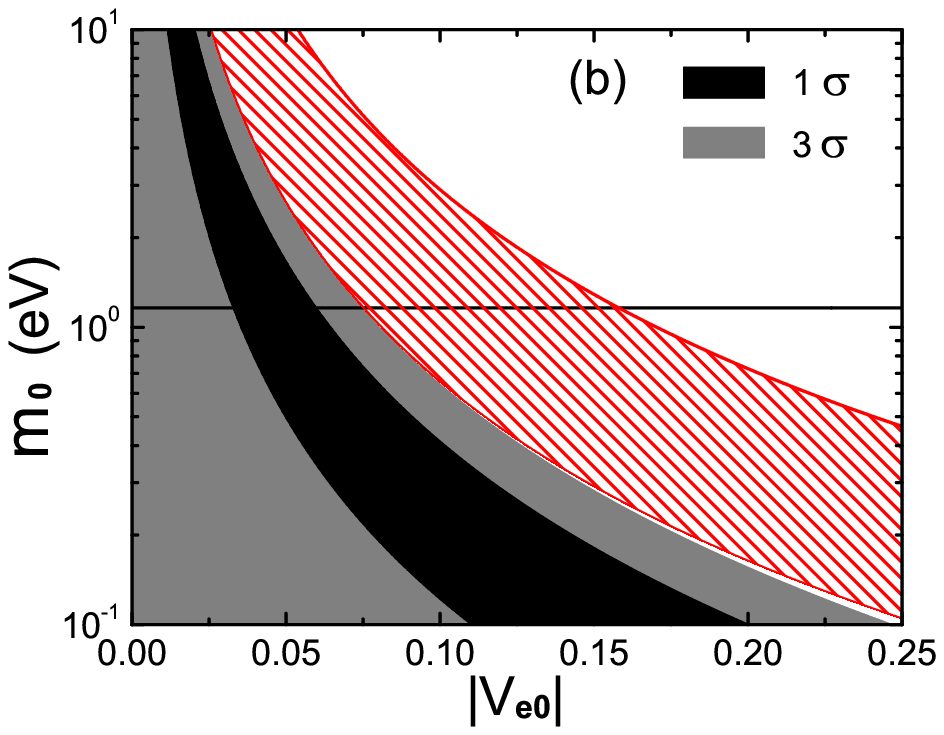}
\end{tabular}
\end{center}
\caption{The regions of ($\rho^{}_{0}$, $\rho^{}_{2}$) (upper panel)
and ($m^{}_{0}$, $|V_{e0}|$) (lower panel) allowed by $\langle
m\rangle_{ee}=0$ and current neutrino oscillation data with $\Delta
m^2_{31}>0$ and $m^{}_{1}=0$. The black and grey scattered regions
(upper panel) and shaded regions (lower panel) stand for $1\sigma$
and $3\sigma$ ranges of the active neutrino data respectively. The
region with (red) sparse lines (lower panel) is constrained from the
sterile neutrino data. The (black) horizontal line stands for the $2
\sigma$ upper bound on the sum of the neutrino masses from
cosmological probes.}
\end{figure}
%%%%%%%%%%%%%%%%%%%%%%%%%%%%%%%%%%%%%%%%%%%%%%%%%%%%%%%%%%

%%%%%%%%%%%%%%%%%%%% Fig. 2 %%%%%%%%%%%%%%%%%%%%%%%%%%%%%%
\begin{figure}[p!]
\begin{center}
\begin{tabular}{cc}
\includegraphics*[bb=25 16 293 236, width=0.70\textwidth]{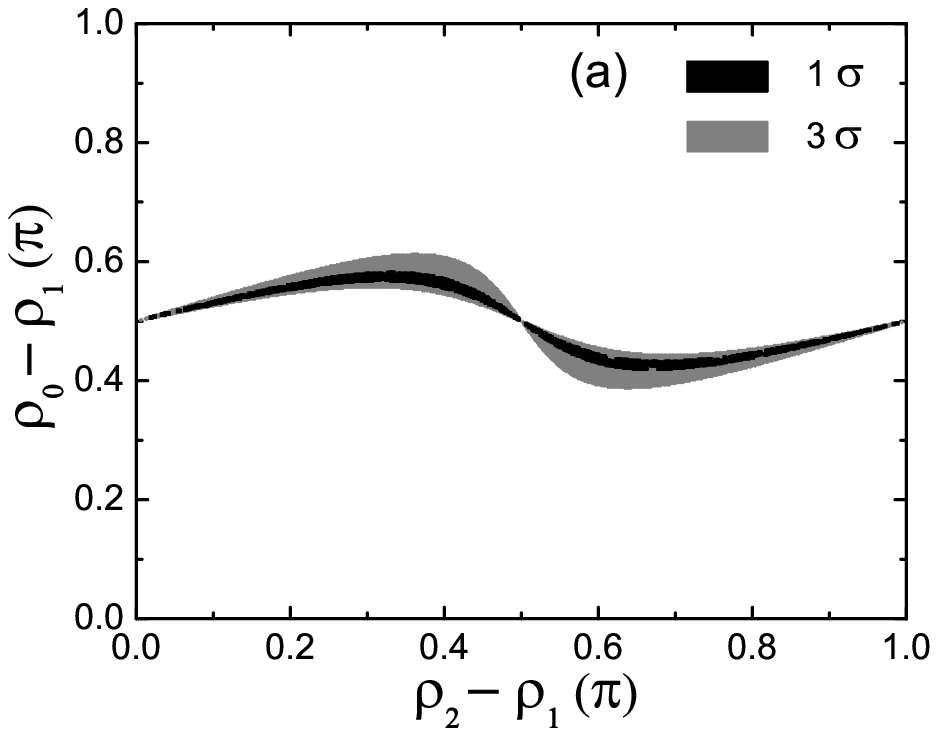}
\\
\includegraphics*[bb=22 20 290 240, width=0.70\textwidth]{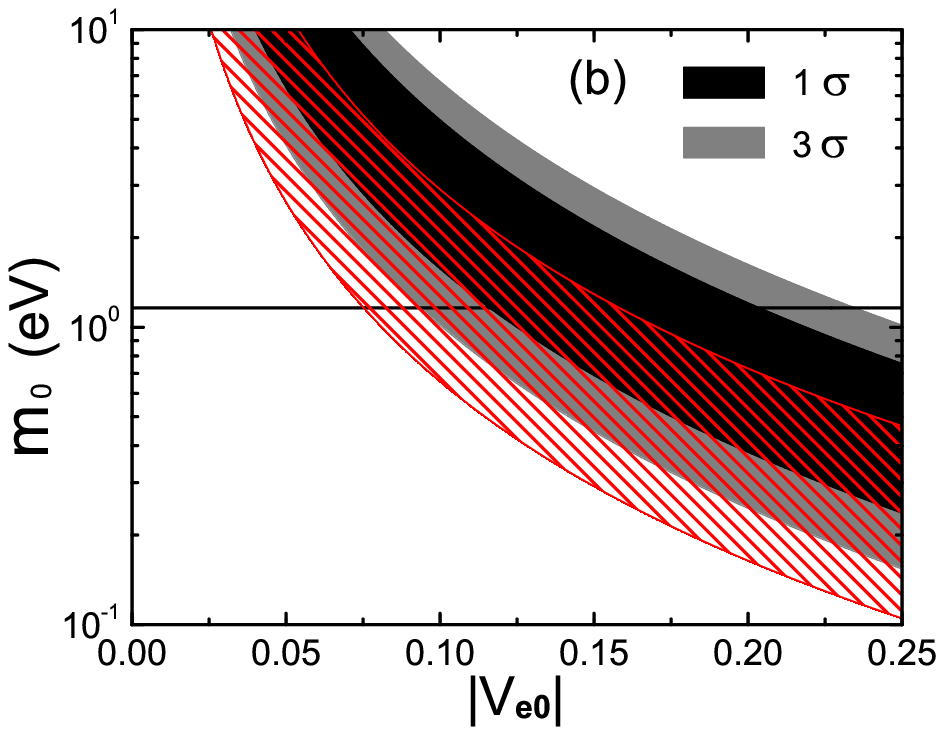}
\end{tabular}
\end{center}
\caption{The regions of ($\rho^{}_{0}-\rho^{}_{1}$,
$\rho^{}_{2}-\rho^{}_{1}$) (upper panel) and ($m^{}_{0}$,
$|V_{e0}|$) (lower panel) allowed by $\langle m\rangle_{ee}=0$ and
current neutrino oscillation data with $\Delta m^2_{31}<0$ and
$m^{}_{3}=0$. The black and grey scattered regions (upper panel) and
shaded regions (lower panel) stand for $1\sigma$ and $3\sigma$
ranges of the active neutrino data respectively. The region with
(red) sparse lines (lower panel) is constrained from the sterile
neutrino data. The (black) horizontal line stands for the $2 \sigma$
upper bound on the sum of the neutrino masses from cosmological
probes.}
\end{figure}
%%%%%%%%%%%%%%%%%%%%%%%%%%%%%%%%%%%%%%%%%%%%%%%%%%%%%%%%%%

%%%%%%%%%%%%%%%%%%%% Fig. 3 %%%%%%%%%%%%%%%%%%%%%%%%%%%%%%
\begin{figure}[p!]
\begin{center}
\begin{tabular}{cc}
\includegraphics*[bb=18 16 298 232, width=0.70\textwidth]{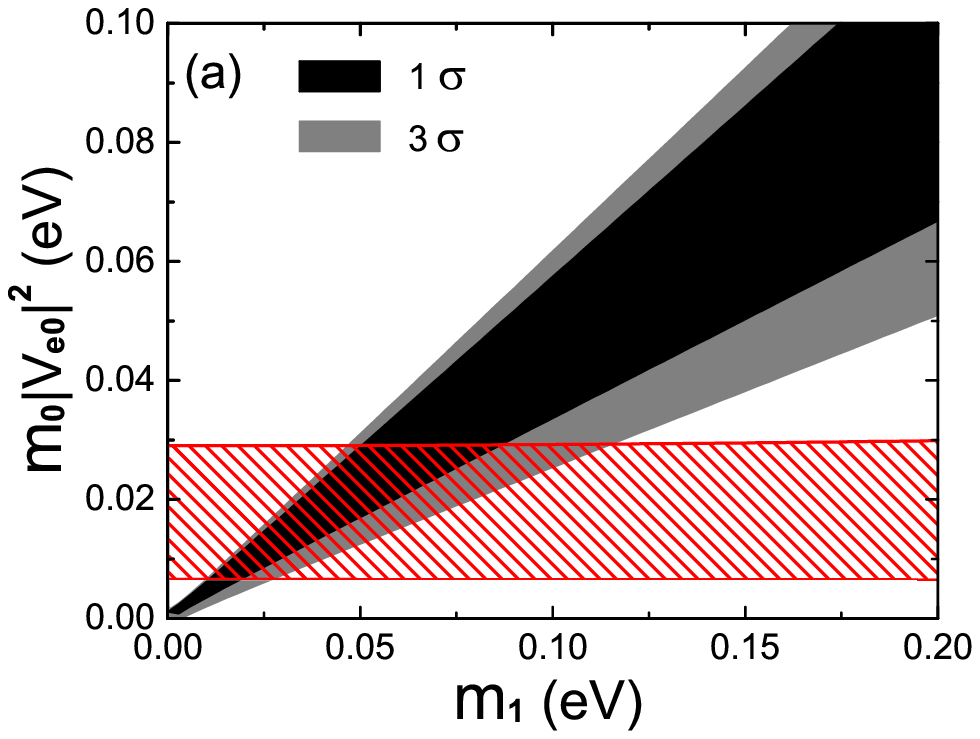}
\\
\includegraphics*[bb=18 20 298 236, width=0.70\textwidth]{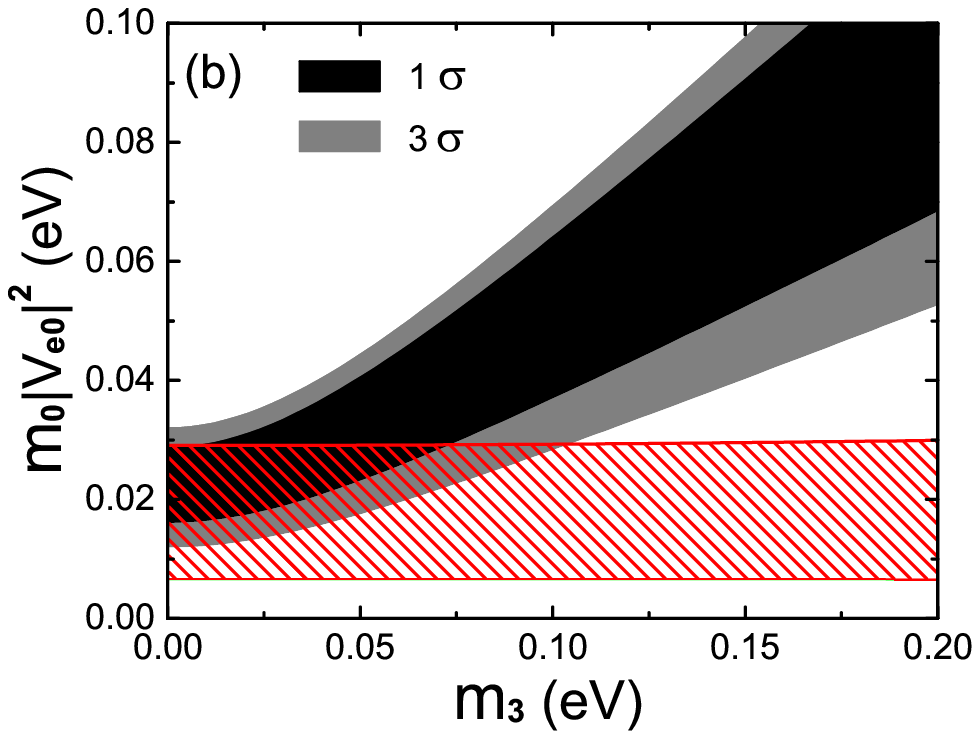}
\end{tabular}
\end{center}
\caption{The regions of $m^{}_{0}|V_{e0}|^{2}$ versus the smallest
neutrino mass ($m^{}_{1}$ or $m^{}_{3}$) with $\Delta m^2_{31}>0$
(upper panel) or $\Delta m^2_{31}<0$ (lower panel). The black and
grey shaded regions stand for $1\sigma$ and $3\sigma$ ranges of the
active neutrino data respectively. The region with (red) sparse
lines is constrained from the sterile neutrino data.}
\end{figure}
%%%%%%%%%%%%%%%%%%%%%%%%%%%%%%%%%%%%%%%%%%%%%%%%%%%%%%%%%%

%%%%%%%%%%%%%%%%%%%% Fig. 4 %%%%%%%%%%%%%%%%%%%%%%%%%%%%%%
\begin{figure}[p!]
\begin{center}
\begin{tabular}{cc}
\includegraphics*[bb=18 17 276 234, width=0.70\textwidth]{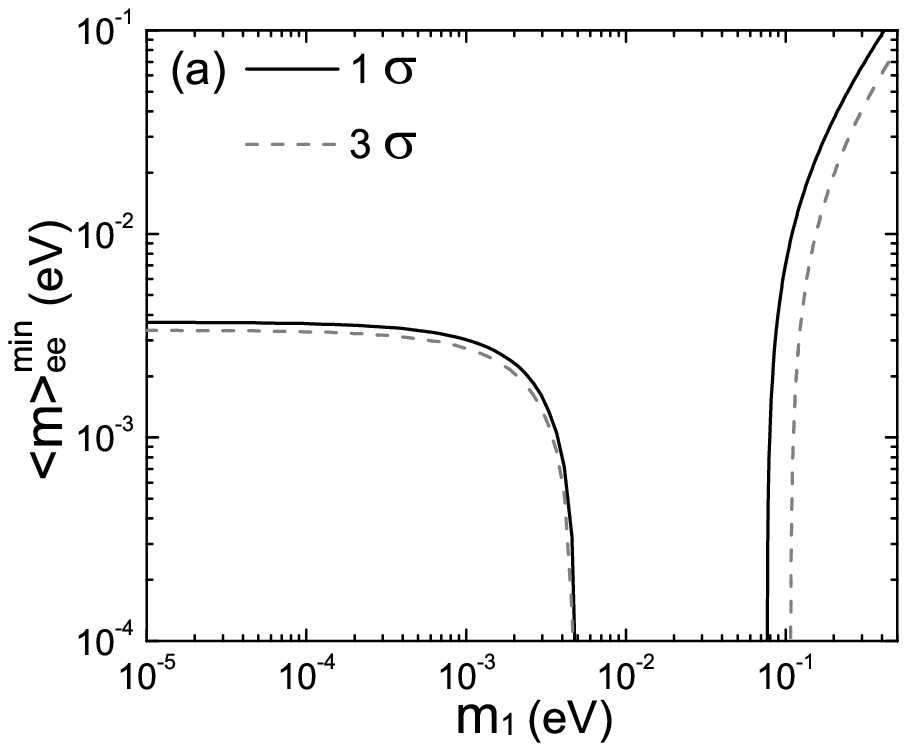}
\\
\includegraphics*[bb=18 17 276 234, width=0.70\textwidth]{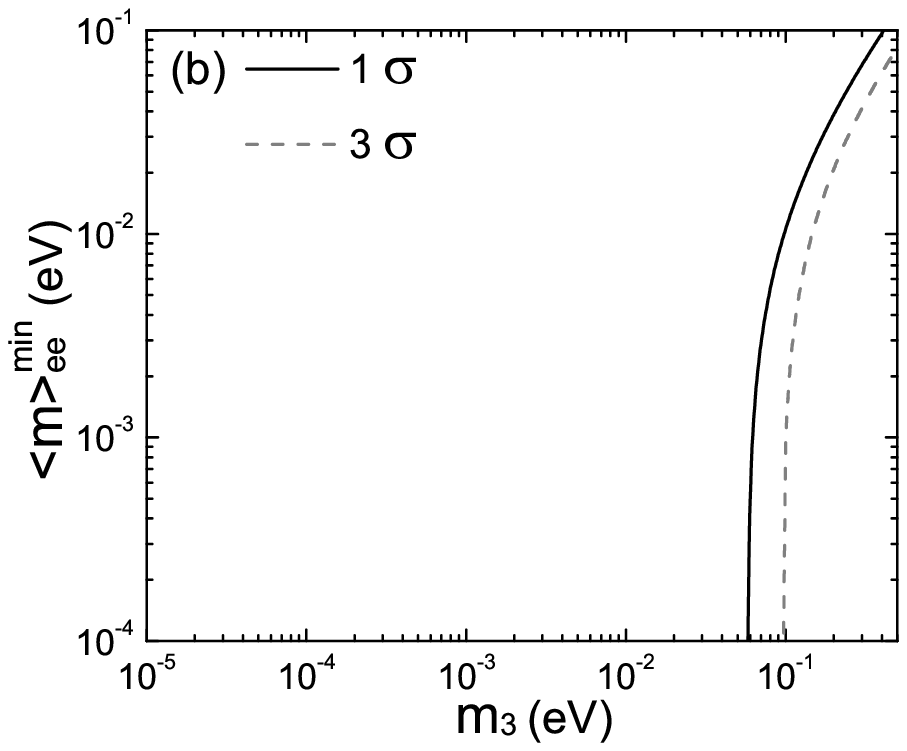}
\end{tabular}
\end{center}
\caption{The minimal values of the effective Majorana mass ($\langle
m\rangle^{min}_{ee}$) versus the the smallest neutrino mass
($m^{}_{1}$ or $m^{}_{3}$) in the (3 + 2) scheme with $\Delta
m^2_{31}>0$ (upper panel) or $\Delta m^2_{31}<0$ (lower panel). The
black solid and grey dashed line stand for $1\sigma$ and $3\sigma$
ranges of the active neutrino data respectively.}
\end{figure}
%%%%%%%%%%%%%%%%%%%%%%%%%%%%%%%%%%%%%%%%%%%%%%%%%%%%%%%%%%


\newpage

\begin{thebibliography}{99}

\bibitem{PDG} Particle Data Group, K. Nakamura {\it et al.},
J. Phys. G {\bf 37}, 075021 (2010).

\bibitem{LSND} A. Aguilar {\it et al.} (LSND Collaboration),
Phys. Rev. D {\bf 64}, 112007 (2001).

\bibitem{Mini} A.A. Aguilar-Arevalo {\it et al.} (MiniBooNE Collaboration),
Phys. Rev. Lett. {\bf 105}, 181801 (2010).

\bibitem{R} G. Mention {\it et al.}, Phys. Rev. D {\bf  83}, 073006
(2011); P. Huber, Phys. Rev. C {\bf 84}, 024617 (2011).

\bibitem{Schwetz4} J. Kopp, M. Maltoni and T. Schwetz,
Phys. Rev. Lett. {\bf 107}, 091801 (2011).

\bibitem{Giunti11a}
C. Giunti and M. Laveder, arXiv:1107.1452 [hep-ph];
arXiv:1109.4033 [hep-ph].

\bibitem{Raffelt}
J. Hamann {\it et al.}, Phys. Rev. Lett. {\bf 105}, 181301 (2010);
E. Giusarma {\it et al.}, Phys. Rev. D {\bf 83}, 115023 (2011);
A.X. Gonzalez-Morales {\it et al.}, arXiv:1106.5052 [astro-ph.CO];
J. Hamann, arXiv:1110.4271 [astro-ph.CO].

\bibitem{Mangano}
G. Mangano and P.D. Serpico, Phys. Lett. B {\bf 701}, 296 (2011);
J. Hamann {\it et al.}, JCAP {\bf 1109}, 034 (2011).

\bibitem{ice}
S. Razzaque and A.Yu. Smirnov JHEP {\bf 1107}, 084 (2011);
R. Gandhi and P. Ghoshal, arXiv:1108.4360 [hep-ph].

\bibitem{Riis}
A.S. Riis and S. Hannestad, JCAP {\bf 1102}, 011 (2011);
J.A. Formaggio and J. Barrett, arXiv:1105.1326 [nucl-ex].

\bibitem{Giunti10}
See, e.g., James Barry {\it et al.}, JHEP {\bf 1107}, 091 (2011); C.
Giunti and M. Laveder, Phys. Rev. D {\bf 82}, 053005 (2010); C.
Giunti, Phys. Rev. D {\bf 61}, 036002 (2000); S.M. Bilenky {\it et
al.}, Phys. Lett. B {\bf 465}, 193 (1999).

\bibitem{review11}
W. Rodejohann, Int. J. Mod. Phys. E {\bf 20}, 1833 (2011).

\bibitem{review03}
S.M. Bilenky {\it et al.}, Phys. Rept. {\bf 379}, 69 (2003).

\bibitem{exps}
See, e.g., H.V. Klapdor-Kleingrothaus {\it et al.}, Eur. Phys. J. A
{\bf 12}, 147 (2001); C.E. Aalseth {\it et al.} [IGEX
Collaboration], Phys. Rev. D {\bf 65}, 092007 (2002); C. Arnaboldi
{\it et al.} [CUORICINO Collaboration], Phys. Rev. C {\bf 78},
035502 (2008).

\bibitem{NME}
See, e.g., A. Faessler {\it et al.}, Phys. Rev. D {\bf 83}, 113015 (2011);
Phys. Rev. D {\bf 79}, 053001 (2009); J. Phys. G {\bf 35}, 075104 (2008).

\bibitem{xing03}
Z.Z. Xing, Phys. Rev. D {\bf 68}, 053002 (2003).

\bibitem{merle}
W. Rodejohann, Nucl.Phys. B {\bf 597}, 110 (2001); A. Merle and W.
Rodejohann, Phys. Rev. D {\bf 73}, 073012 (2006); S. Dev and S.L.
Kumar, Mod. Phys. Lett. A {\bf 22}, 1401 (2007); J. Jenkins, Phys.
Rev. D {\bf 79}, 113003 (2009); Y. BenTov and A. Zee, Phys.Rev. D
{\bf 84}, 073012 (2011).

\bibitem{Fogli}
G.L. Fogli {\it et al.}, Phys. Rev. D {\bf 84}, 053007 (2011).

\bibitem{Schwetz3}
T. Schwetz {\it et al.}, New J. Phys. {\bf 13}, 109401 (2011).

\bibitem{WMAP10}
E. Komatsu {\it et al.} (WMAP Collaboration), Astrophys. J. Supp.
{\bf 192}, 18 (2011).

\bibitem{cosmosnu}
Y.Y.Y. Wong, Ann. Rev. Nucl. Part. Sci. {\bf 61}, 69 (2011); M.C.
Gonzalez-Garcia {\it et al.}, JHEP {\bf 1008}, 117 (2010).

\bibitem{Probe}
F. Simkovic {\it et al.} Phys. Rev. D {\bf 82}, 113015 (2010); G.L.
Fogli {\it et al.}, Phys. Rev. D {\bf 80}, 015024 (2009); F.
Deppisch and H. Pas, Phys. Rev. Lett. {\bf 98}, 232501 (2007).

\bibitem{xing09}
Z.Z. Xing, Phys. Lett. B {\bf 679}, 255 (2009); arXiv:1110.0083
[hep-ph].

\bibitem{RHN}
See, e.g., P. Benes {\it et al.}, Phys. Rev. D {\bf 71}, 077901
(2005); A.de Gouvea {\it et al.}, Phys. Rev. D {\bf 75}, 013003
(2007); W. Rodejohann, Phys. Lett. B {\bf 684}, 40 (2010); M.
Blennow {\it et al.}, JHEP {\bf 1007}, 096 (2010); M. Mitra {\it et
al.}, arXiv:1108.0004 [hep-ph].

\bibitem{LR}
C.S. Aulakh and R.N. Mohapatra, Phys. Lett. B {\bf 119}, 136 (1982);
R.N. Mohapatra, Phys. Rev. D {\bf 34} 909 (1986); J.D. Vergados,
Phys. Lett. B {\bf 184}, 55 (1987).


\bibitem{Futu11}
H. Ejiri {\it et al.}, Phys. Rev. Lett. {\bf 85}, 2917 (2000); K.
Zuber, Phys. Lett. B {\bf 519}, 1 (2001); C. Arnaboldi {\it et al.}
[CUORE Collaboration], Astropart. Phys. {\bf 20}, 91 (2003); R.
Arnold {\it et al.} [SuperNEMO Collaboration], Eur. Phys. J. C {\bf
70}, 927 (2010).

\bibitem{Futu22}
A. Terashima {\it et al.} [KamLAND Collaboration], J. Phys. Conf.
Ser. {\bf 120} 052029 (2008); R. Gaitskell {\it et al.} [Majorana
Collaboration], arXiv:nucl-ex/0311013; M.C. Chen [SNO+
Collaboration], arXiv:0810.3694 [hep-ex].

\end{thebibliography}
\end{document}